\documentclass[a4paper,12pt]{article}
\usepackage{amsfonts,amsthm,amsmath,amssymb,graphicx,hyperref,color,youngtab}

\def\Tr{{\rm Tr\, }}
\newcommand{\be}{\begin{equation}}
\newcommand{\bea}{\begin{eqnarray}}
\newcommand{\ee}{\end{equation}}
\newcommand{\eea}{\end{eqnarray}}

\begin{document}

\makeatletter
\@addtoreset{equation}{section}
\makeatother
\renewcommand{\theequation}{\thesection.\arabic{equation}}
\vspace{1.8truecm}
%%%%%%%%%%%%%%%%%

{\LARGE{ \centerline{\bf Scrambling in Yang-Mills}  }}  

\vskip.5cm 

\thispagestyle{empty} 
\centerline{ {\large\bf Robert de Mello Koch$^{a,b,}$\footnote{{\tt robert@neo.phys.wits.ac.za}},
Eunice Gandote$^{b,}$\footnote{{\tt eunice@aims.edu.gh}}  }}
\centerline{{\large\bf and
Augustine Larweh Mahu${}^{b,c,}$\footnote{ {\tt aglarweh@gmail.com}} }}

\vspace{.4cm}
\centerline{{\it ${}^{a}$ School of Physics and Telecommunication Engineering},}
\centerline{{ \it South China Normal University, Guangzhou 510006, China}}

\vspace{.4cm}
\centerline{{\it ${}^{b}$ National Institute for Theoretical Physics,}}
\centerline{{\it School of Physics and Mandelstam Institute for Theoretical Physics,}}
\centerline{{\it University of the Witwatersrand, Wits, 2050, }}
\centerline{{\it South Africa }}

\vspace{.4cm}
\centerline{{\it ${}^{c}$ Department of Mathematics,,}}
\centerline{{\it  University of Ghana, P. O. Box LG 62, Legon, Accra, Ghana.}}

\vspace{1truecm}

%%%%%%%%%%%%%%%%%
\thispagestyle{empty}

\centerline{\bf ABSTRACT}

\vskip.2cm 
Acting on operators with a bare dimension $\Delta\sim N^2$ the dilatation operator of $U(N)$ ${\cal N}=4$ super Yang-Mills 
theory defines a 2-local Hamiltonian acting on a graph.
Degrees of freedom are associated with the vertices of the graph while edges correspond to terms in the Hamiltonian.
The graph has $p\sim N$ vertices.
Using this Hamiltonian, we study scrambling and equilibration in the large $N$ Yang-Mills theory.
We characterize the typical graph and thus the typical Hamiltonian.
For the typical graph, the dynamics leads to scrambling in a time consistent with the fast scrambling conjecture.
Further, the system exhibits a notion of equilibration with a relaxation time, at weak coupling, given by 
$t\sim{p\over\lambda}$ with $\lambda$ the 't Hooft coupling.

\setcounter{page}{0}
\setcounter{tocdepth}{2}
\newpage
\tableofcontents
\setcounter{footnote}{0}
\linespread{1.1}
\parskip 4pt

{}~
{}~

\section{Introduction}

Black holes in general relativity exhibit incredibly fast relaxation time scales.
Since the AdS/CFT correspondence claims an equivalence between conformal field theories in $d$ dimensions and theories 
of quantum gravity on negatively curved spacetimes\cite{Maldacena:1997re,Witten:1998qj,Gubser:1998bc}, the mechanism
behind these extremely rapid thermalization rates should be coded into the dynamics of large $N$ and strongly coupled 
conformal field theories.
Motivated by this issue we study scrambling and equilibration in ${\cal N}=4$ super Yang-Mills theory, with gauge group $U(N)$.
There are at least two features of our study that must be improved before we can make contact with the 
physics of black holes.
First, operators in the conformal field theory corresponding to a black hole necessarily have a 
very large dimension $\Delta\sim N^2$.
The generic operator is constructed using the complete collection of fields in the theory.
Although our operators have a dimension of order $N^2$, they are special in that they are constructed 
using three complex adjoint scalars and two complex adjoint fermions.
Second, the link to classical gravity emerges in the strong coupling limit of the field theory.
Our analysis is limited to weak coupling.
However, we will see that our simplified system is already interesting.

Recall that the AdS/CFT correspondence identifies the dimensions of operators in the conformal field 
theory with the energies of energy eigenstates in the dual gravitational theory.
This has been pursued in exquisite detail in the planar limit of ${\cal N}=4$ super Yang-Mills 
theory\cite{Beisert:2010jr}, where the identification of the dilatation operator $D$ with a Hamiltonian is particularly 
fruitful because $D$ is the Hamiltonian of an integrable spin chain.
The energy of a spin chain state equals the dimension of the corresponding operator.
The dynamics of the worldsheet string theory is also integrable \cite{Bena:2003wd} and there is an exact match
between string theory energies and operator dimensions \cite{Kazakov:2004qf}.
Although integrability allows us to go beyond weak coupling, the planar limit is not the correct arena for the questions 
we consider.
Indeed, integrable systems do not thermalize in the conventional way: they do not thermalize to a Gibbs ensemble. 
Integrable systems thermalize into a ``generalised Gibbs ensemble" due to the existence of many extensive 
conserved charges. 
This is well understood for integrable systems relaxing after a quantum quench\cite{GGE}. 
Further, completely integrable models can never exhibit chaos, but the holographic dual to a black hole is expected to 
exhibit chaotic dynamics \cite{Shenker:2013pqa}.

An interesting extension beyond the planar limit considers operators whose bare dimension grows 
parametrically with $N$ as we take $N\to\infty$.
The mixing problem of these heavy operators has new complications absent in the planar limit: single trace operators can
and do mix so multi trace structures must be included in the problem and they all mix in a non-trivial way.
A second complication is that the sheer number of non-planar diagrams is so big that it overcomes the
usual higher genus suppression and we must sum more than just the planar 
diagrams \cite{Balasubramanian:2001nh,Aharony:2002nd,Berenstein:2003ah}.
The final complication arises because as the number of fields in the multi trace operator grows beyond $N$
there are trace relations which express the equality of naively distinct multi trace structures\footnote{For example, 
invariants of a single matrix are written in terms of the eigenvalues of the matrix.
Given $N$ independent invariants, the eigenvalues and hence all invariants are determined.
As a consequence, there are relations between invariants expressed as a collection of terms that sum to zero.
Each term is of a fixed degree in the matrix and different terms have different trace structures. 
An example of a relation of this type is provided by the Cayley-Hamilton Theorem and by the Mandelstam 
relations \cite{Berenstein:1993gb}.}.  
Starting with \cite{Corley:2001zk} methods based on group representation theory were employed to address all 
three of these issues in a single complex matrix model.
A linear basis for multi-matrix invariants, the restricted Schur polynomials, which we use in this work, is constructed in 
\cite{Bhattacharyya:2008rb,Bhattacharyya:2008xy} (see also \cite{Balasubramanian:2004nb}).
Although we will not use them in our study, note that closely related bases were introduced and studied 
in \cite{Kimura:2007wy,Brown:2007xh,Brown:2008ij,Kimura:2008ac}.
The restricted Schur polynomials are labeled by a collection of Young diagrams, one for each species of field appearing in 
the operator, plus one more denoted $R$ for the complete collection of fields. 
They diagonalize the free field theory two point function, explicitly take all finite $N$ trace relations into account and mix
only weakly at one loop.
Summing the complete set of ribbon graphs contributing to a free field  theory correlator is reduced to rather straight 
forward manipulations in group theory: the computation of projection operators and matrices representing permutations, 
as well as commutators, products and traces of them.

Our focus is on operators constructed using $O(N^2)$ fields.
The majority of the fields appearing in the operators we study are a single complex adjoint scalar (say $\phi_1$).
There are a smaller number of additional scalar ($\phi_2$ and $\phi_3$) as well as fermion ($\psi_1$ and $\psi_2$) fields, all 
transforming in the adjoint of $U(N)$.
We will use $n_i$ to denote the number of $\phi_i$ fields and $m_i$ to denote the number of $\psi_i$ fields.
In the limit where the row lengths of the Young diagram $R$ labeling the restricted Schur polynomials are all different, with the
difference $\gg1$ (called the displaced corners approximation \cite{Carlson:2011hy,Koch:2011hb} because the corners on the
right hand side of the Young diagram are well separated) the mixing problem simplifies dramatically.
New symmetries appear and these naturally suggest that the state space can be labeled with a pair of Young diagrams (describing
the $\phi_1$ fields and one more, denoted $R$, for the complete collection of fields) and a graph for the remaining fields
\cite{Koch:2011hb,deMelloKoch:2012ck}.
The mixing problem can be diagonalized on the Young diagram labels, leaving a Hamiltonian describing dynamics on a 
graph\cite{deCarvalho:2018xwx,deMelloKoch:2020agz}.
Vertices of the graph correspond to rows (for a short and wide diagram) or columns (for a tall and thin diagram)
of the Young diagram $R$ and hence they correspond to dual giant and giant graviton branes.
As a consequence of the displaced corners condition the branes are separated in spacetime. 
Edges stretching between vertices correspond to open strings that stretch between branes.
In a suitable adiabatic limit, reviewed in Appendix \ref{fromD}, these modes are frozen, i.e. they do not evolve in time.
Each brane can be excited, which is represented as a closed loop made out of a single edge attached to a given vertex.
In the adiabatic limit these excitations of a particular brane are the only dynamical degrees of freedom.
These degrees of freedom live at the vertices of the graph and they are able to hop to any other site as long as there is an edge
in the graph that connects the two sites \cite{deCarvalho:2018xwx,deMelloKoch:2020agz}.

Thus, the spin chain of the planar mixing problem is replaced by dynamics on a graph, when the mixing problem of
heavy operators is considered.
It is noteworthy that dynamics on a graph naturally emerges in this way.
Indeed, models describing the dynamics on graphs were used to examine the fast scrambling conjecture\cite{Sekino:2008he},
first in \cite{Lashkari:2011yi}, which was followed by a number of interesting 
articles\cite{Bentsen:2018uph,Hartmann:2019cxq,Lucas:2019aif,Chen:2019klo,Garcia-Garcia:2020cdo,Xu:2020shn}\footnote{These studies 
use an ``interaction graph''. 
Degrees of freedom live at the vertices of the interaction graph. 
The interaction graph has an edge between two vertices if and only if the Hamiltonian includes an interaction term 
for these degrees of freedom. 
There is a simple relation between the graph that emerges from Yang-Mills theory, called a Gauss graph in 
\cite{deMelloKoch:2012ck} and the interaction graph: dropping the closed loops from the Gauss
graph one obtains the interaction graph. We stick to this terminology in this article.}.
The logic of \cite{Lashkari:2011yi} is elegant and worth summarizing.
Consider a state of some subsystem $S$, and denote the complementary subsystem to $S$ by $S^c$. 
By saying that information is scrambled we mean it is hidden in complicated correlations between subsystems $S$ and $S^c$.
Using this observation one can argue that scrambling subsystem $S$ is the same as signaling to $S^c$.
Thus, bounds on signaling are immediately bounds on scrambling.
With this insight, \cite{Lashkari:2011yi} appeals to classic methods of Lieb-Robinson\cite{Lieb:1972wy} which bound
signaling by proving bounds on commutators $[O_A(t),O_B]$ where $O_A$ and $O_B$ are observables localized on 
disjoint subsystems $A$ and $B$ of a lattice spin system.
In this way \cite{Lashkari:2011yi} bound the signaling time for Hamiltonians with dense two body interactions\footnote{Dense
means the number of interacting pairs of degrees of freedom scales like $n^2$.} to no faster than $O(\log n)$ with $n$ the
number of degrees of freedom.
The resulting bound refers to the maximum degree $D_V$ of any vertex of the interaction graph.
$D_V$ also appears in the assumption that each term in the Hamiltonian is bounded by $c/D_V$ with $c$ some constant that
does not scale with the size of the system.
The Lieb-Robinson bound then says that a suitably normalized commutator is bounded by $\sim{1\over D_V}e^{8ct}$.
Using this bound, its now possible to show that for times $\sim \log D_V$ the reduced density matrix on each site $i$ 
is approximately a pure state.
$D_V$ is the maximum vertex degree, so we expect $D_V\sim n$.
Since scrambling requires entanglement,  this bounds the scrambling time to be at least $\sim \log n$.

In this paper we study the dynamics of the Hamiltonian defined by the mixing problem for heavy operators, described 
by dynamics on a graph.
The relevant Hamiltonian is described in Section \ref{dynamics}.
Our Hamiltonian describes the physics of bound states of giant gravitons and their excitations.
The number of giants in the boundstate is large enough to backreact and produce a new spacetime geometry\cite{Lin:2004nb}.
By choosing the right boundstate of giant gravitons excited in a particular way, we would produce operators dual to
black holes.
A black hole state would have a number of general features that one could look for\footnote{For a very readable and informative
discussion we recommend \cite{Balasubramanian:2005mg,Balasubramanian:2007bs,Fareghbal:2008ar}.}.
First, the mass of the black hole in AdS translates, upon using the standard AdS/CFT dictionary, into a scaling dimension 
for operators that grows as $\Delta\sim N^2$.
To explain the entropy of the black hole, the number of operators should be $\sim e^{b N^2}$ with $b$ some constant
that does not depend on $N$.
We verify these expectations in Section \ref{graphs}.
Each operator is labeled by a different graph and hence by a different Hamiltonian.
By numerically generating the complete set of graphs for finite values of $N$ (where numerical analysis is still
possible), we give evidence that there is a ``typical'' graph and that almost every graph, at large $N$, looks like
the typical graph.
This typical graph defines a typical Hamiltonian and it is the dynamics of this typical Hamiltonian that we consider.
In Section \ref{scramble} we study scrambling, establishing a Lieb-Robinson bound which ensures that the system
does not scramble faster the bound implied by the fast scrambling conjecture.
We also explore entanglement generation for the typical dynamics.
The leads to a puzzle: the recurrence time is much smaller than we expect.
In Section \ref{equal} we show for a conveniently chosen initial non-equilibrium state, that the system
evolves to thermal equilibrium and we estimate the thermalization time scale.
The puzzle of the recurrence time is also resolved: we argue that as far as the dynamics is concerned, the typical Hamiltonian 
is rather special and does not give a reliable description of the physics.
Small fluctuations in the typical Hamiltonian are important and must be included.
In Section \ref{discuss} we discuss our results and outline some future directions.

\section{Dynamics on Gauss graphs}\label{dynamics}

As reviewed in Appendix \ref{fromD} the dynamics we consider is of a system of bosons, hopping on a lattice.
The lattice is defined by a directed graph $G=(V,E)$, where $V$ is a set of vertices and $E$ a set of directed edges.
In what follows we always use $p=|V|$ to denote the total number of vertices. 
Since the edges are directed it makes sense to talk about edges departing from a vertex or edges arriving at a vertex.
Not just any directed graph is allowed; at each vertex the number of arriving edges must equal the number of departing edges.

The bosons live at the vertices of the graph.
Thus at each vertex $i\in V$ we have a bosonic Fock space ${\cal F}_i$.
The full Fock space is a tensor product ${\cal F}\equiv {\cal F}_1\otimes {\cal F}_2\otimes\cdots\otimes{\cal F}_p$.
Associated to the $i$th Fock space is a pair of oscillators $b_i,b_i^\dagger$ and a vacuum state $|0\rangle_i$.
The oscillators $b_i,b_i^\dagger$ act as the identity on all ${\cal F}_j$ with $j\ne i$, and in the usual way on ${\cal F}_i$.
The algebra of the bosonic operators is
\bea
[b_i,b_j^\dagger]=\delta_{ij}
\eea
The $i$th Fock space vacuum obeys $b_i |0\rangle_i =0$ and the vacuum of the full Fock space ${\cal F}$ is given by
\bea
|0\rangle = |0\rangle_1\otimes |0\rangle_2 \otimes\cdots\otimes |0\rangle_p
\eea

To write the Hamiltonian describing the dynamics of these bosons, it is useful to introduce the $p\times p$ matrix $N_{ij}$.
The matrix elements $N_{ij}$ count how many edges stretch between vertices $i$ and $j$, regardless of orientation.
As far as the Hamiltonian is concerned, we can ignore the orientation of edges which corresponds to treating $G$ as an
undirected graph.
Thus, $N_{ij}$ is a symmetric matrix with zeros on the diagonal, that completely determines the graph $G$.
As an example, consider the following graph
\bea
\begin{gathered}\includegraphics[scale=0.1]{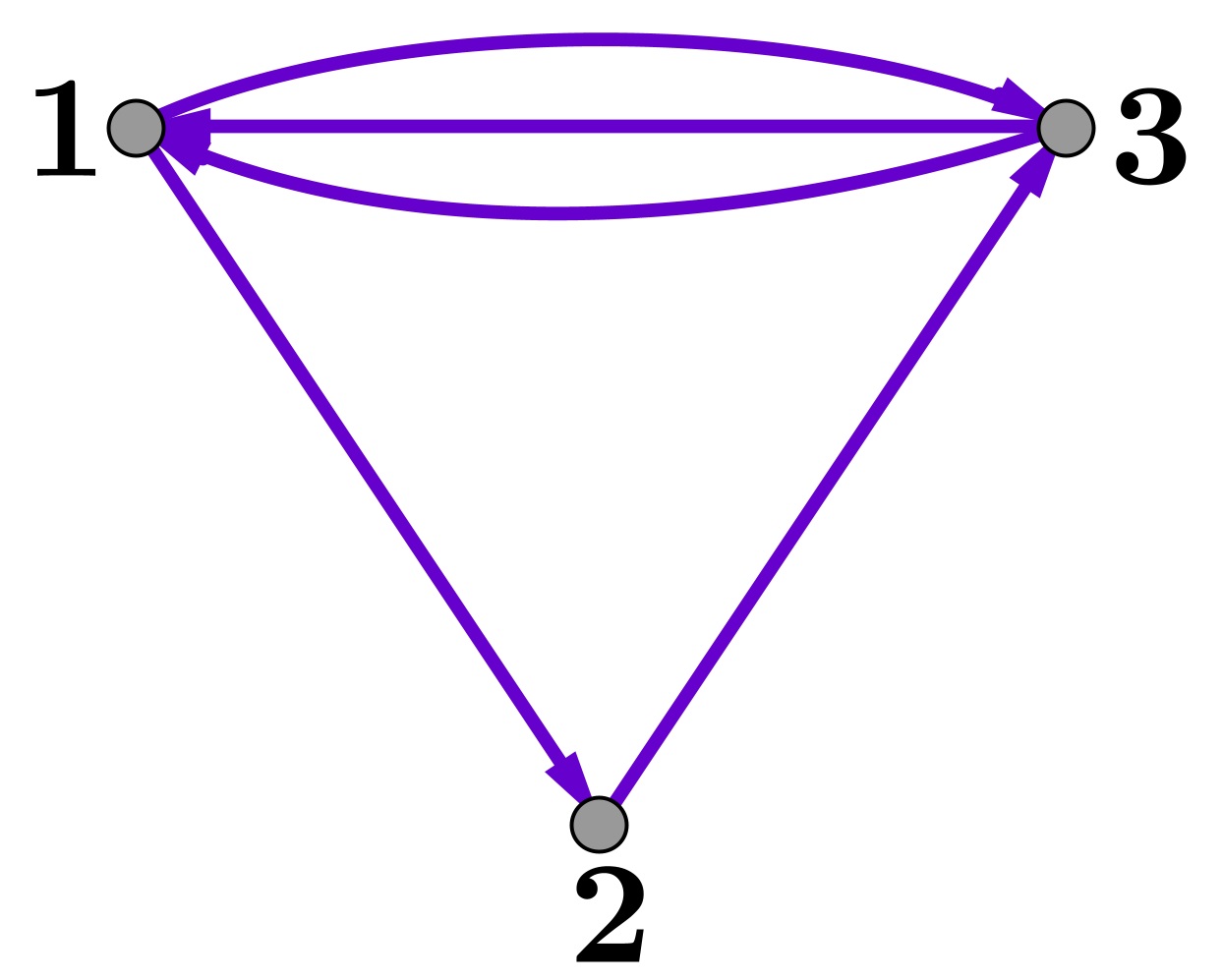}\end{gathered}\qquad
N_{ij}=\left[\begin{matrix}0 &1 &3\\ 1 &0 &1\\ 3 &1 &0\end{matrix}\right]
\eea
In terms of this matrix, the Hamiltonian we study is given by
\bea
H&=&{g_{YM}^2\over (4\pi)^2}\sum_{i,j=1,i\ne j}^p (\sqrt{r_i}-\sqrt{r_j})^2 N_{ij}
+{2g_{YM}^2\over (4\pi)^2}\sum_{i=1}^p\, {r_i \over l_i}k_{i}\, b^\dagger_{i} b_{i}
-{2g_{YM}^2\over (4\pi)^2}\sum_{i,j=1,i\ne j}^p\sqrt{r_i r_j\over l_{i}l_{j}} N_{ji}b^\dagger_{j} b_{i}\cr
&&\label{GGham}
\eea
where
\bea
r_i=N+l_{i}\qquad k_{i}=\sum_{l=1,l\ne i}^p (N)_{il}
\eea
The parameters of the model are $N$, $l_{i}$, $g_{YM}^2$, $p$ and the matrix $N_{ij}$.
$N$ sets the rank of the gauge group of the Yang-Mills theory and $g_{YM}^2$ is the coupling constant.
We study the large $N$ limit, at weak 't Hooft coupling.
The parameters $l_i$, $i=1,2,\cdots,p$ are positive integers of order $\sim N$.
In the CFT they set the row lengths of Young diagram $R$ labeling our operator.
They are ordered so that $l_i>l_j$ if $j>i$.
The displaced corners approximation requires that $|l_i-l_j|\gg 1$ for all $i\ne j$.
In the holographic dual $l_i$ is the angular momentum of the corresponding dual giant graviton.
We are interested in the limit in which $p$ goes to infinity.
If we take $p=\epsilon N$ with $\epsilon\ll 1$, we can consider operators with differences in lengths of adjacent rows of $R$
of order $\sim \epsilon^{-1}\gg 1$ which justifies the displaced corners approximation.
In the next section we study the graphs relevant for our problem, thereby characterizing the matrices $N_{ij}$.
We focus on graphs with number of edges $|E|\approx p^2 =\epsilon^2 N^2$. 
In this case, the bare dimension of our operator is $\Delta\sim \epsilon N^2$ and the number of fields scale as
$n_1\sim \epsilon N^2$ and $n_2\sim n_3\sim m_1\sim m_2\sim \epsilon^2 N^2$ as we take $N\to\infty$.

The spectrum of the Hamiltonian has an interesting structure.
The first term in the Hamiltonian is an order $\sim 1$ number times the 't Hooft coupling $\lambda =g_{YM}^2 N$.
This term is a constant, determined by the number of edges and the specific vertices the edges stretch between.
The second term is a constant, equal to the total number of bosons hopping in the graph.
Since the Hamiltonian preserves particle number we can restrict the dynamics 
to a subspace with fixed total number of particles.
We work on the subspace with $N_b$ bosons hopping on the graph.
These first two terms give the largest contribution to the energy eigenvalues.
The remaining terms give a much smaller correction to the first two terms.
These small corrections resolve the degeneracies of the multiparticle Fock space.
In Section \ref{LRBound} we estimate the size of the terms in the Hamiltonian.
The first two terms are of size $\sim\lambda$, and that the remaining terms are of size $\sim\epsilon^2\lambda$.
The dimension of the multi particle Fock space grows very rapidly: for $N_b$ bosons hopping on a graph with $p$ vertices
the dimension of the relevant subspace of Fock space is given by
\bea
dim_{p,N_b}={(p+N_b-1)!\over N_b! (p-1)!}
\eea
The Hamiltonian we consider computes the one loop anomalous dimension $E_1$, which corrects the bare dimension,
itself of order $E_0=\epsilon N^2$.
The pattern for the possible $E_1$ values present in the above spectrum, is a set of levels separated by gaps of 
order $\sim \lambda$, with each level a collection of an enormous numbers of nearly degenerate states, with splitting 
$\sim \epsilon^2 \lambda$.
Using a measuring apparatus that can resolve energy differences $\sim N^2$, but not the much smaller scales $\sim\lambda$
or $\sim \epsilon^2 \lambda$ we would only resolve a coarse grained version of the physics. 
After coarse graining its not possible to distinguish between these almost degenerate states, so we
naturally obtain macrostates with a large entropy.
This is a promising start to explain the black hole entropy. 
One check of this idea is to count the total number of operators that can be defined.
Since there is an operator associated to every graph (see Appendix \ref{fromD}) the number of graphs should be large
enough ($\sim e^{bN^2}$) for this idea to work.

An important technical comment is in order: the studies of the fast scrambling conjecture given 
in \cite{Lashkari:2011yi,Bentsen:2018uph,Lucas:2019aif,Chen:2019klo} which were an important motivation for this study,
make use of the assumption that the Hamiltonian (and other operators) have a finite norm.
The Hamiltonian defined in (\ref{GGham}), is unbounded.
Thus, it seems that the methods of finite dimensional quantum mechanics can not be used and a careful treatment of the 
system with the methods of functional analysis \cite{Reed:1975uy} is necessary. 
This conclusion is too hasty and too pessimistic.
The Hamiltonian in (\ref{GGham}) conserves particle number.
Thus, if we restrict to initial states with finite particle number, the whole evolution happens in a finite dimensional subspace 
of the Fock space. 
In this case the Hamiltonian $H$ and all relevant observables can be represented by bounded operators on this subspace
so that we are back in the framework of finite dimensional quantum mechanics\cite{equilibration}.

Our Hamiltonian is derived by evaluating the action of the dilatation operator on a specific class of heavy operators in
${\cal N}=4$ super Yang-Mills.
It is interesting to note that a closely related model was suggested and studied in \cite{Magan:2016ojb} as a toy model of
black hole dynamics.
See also \cite{Iizuka:2008hg,Iizuka:2008eb} for related work.

\section{Properties of Gauss Graphs}\label{graphs}

The graphs arising from the operator mixing problem of Yang-Mills theory were called Gauss graphs 
in \cite{deMelloKoch:2012ck}.
Gauss graphs are graphs with directed edges and any number of vertices.
In addition, at every vertex in the graph, the number of edges terminating on the vertex is 
equal to the number of edges departing from the vertex.
We call this the Gauss constraint. 
By removing edges that have both endpoints at a single vertex (so these edges form a closed loop) we obtain the interaction
graph.
Edges of the interaction graph are always stretched between distinct vertices.
A directed graph obeying the Gauss constraint is called a {\it balanced directed graph} \cite{GraphWiki} in the mathematics
literature.
In this section we describe an algorithm that can be used to generate the complete set of interaction graphs, given that
each graph has $p$ vertices and $E$ edges.
The number of interaction graphs grows extremely rapidly so that is makes sense to talk about the ``typical graph''.
We characterize properties of the typical graph, using numerical results. 
For each interaction graph there is a Hamiltonian.
By characterizing the typical graph we are characterizing the typical Hamiltonian.
We can then study the scrambling time and relaxation rates of this typical Hamiltonian.

\subsection{Generating interaction graphs}

The key difficulty in generating interaction graphs entails respecting the Gauss constraint.
Consider some interaction graph $G$.
Our first observation is that any closed oriented path, made from edges belonging to $G$, respects the Gauss constraint.
Deleting the edges that make up this path produces a new graph $G'$, which itself also obeys the Gauss constraint, i.e. $G'$
is also an interaction graph.
We can now repeat the procedure: construct any closed path, made from edges belonging to $G'$.
Delete this new path to find a new interaction graph $G''$.
This procedure can be repeated until all edges in $G$ have been deleted, and so $G$ has been decomposed into a collection
of closed paths.
To generate the interaction graph $G$ we follow the reverse process in which we ``grow'' $G$ by dressing a bare set of 
vertices with closed oriented paths. 

Its easy to understand why this decomposition is always possible: choose any given edge in the graph and consider
the vertex that this edge ends on. 
The Gauss constraint guarantees that there is always an edge leaving this vertex, that can be joined with the edge we have
to produce the second edge in the path.
We can keep growing the path in this way.
The growing process terminates when the last edge we consider can be joined with the first edge in the path, producing
a closed path.
The point is that the Gauss constraint implies that any edges left after a closed path is deleted, belong to a closed path 
and hence as long as there are edges left, we can keep making closed paths.

\begin{figure}[h]
     \centering
     \includegraphics[width=0.8\textwidth]{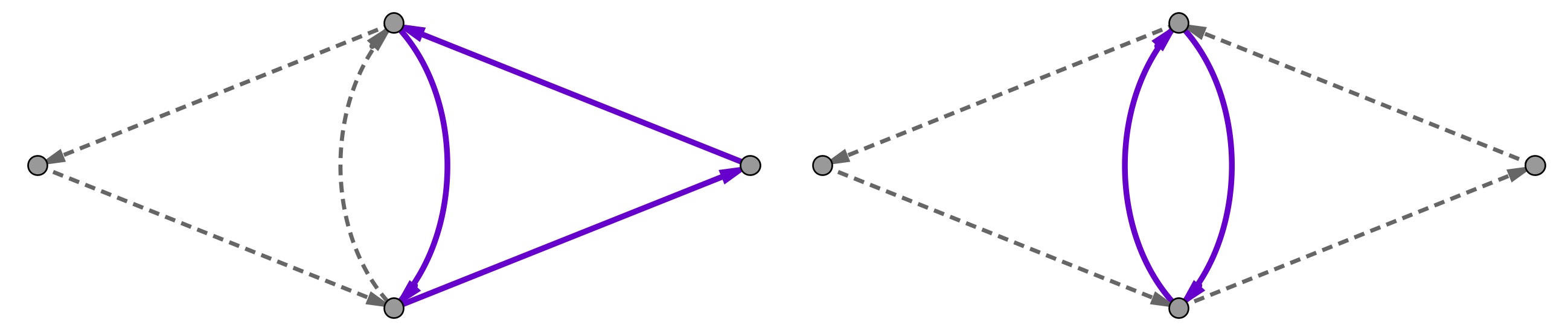}
     \caption{The interaction graph shown can be decomposed into two paths, each of length 3 as shown on the left, or into two
                   paths, one of length 2 and one of length 4 as shown on the right.}
     \label{fig:GGdecomp}
\end{figure}
The decomposition of an interaction graph into closed paths is not unique.
Indeed, consider the example shown in Figure \ref{fig:GGdecomp}.
The interaction graph shown, with a total of 6 edges, can be decomposed into two paths of length 3, or into one path
of length 4 and one path of length 2.

\begin{table}[h!]
\centering
{\footnotesize
\begin{tabular}{|| c || c | c | c || c | c | c ||c | c | c ||}
%\hline 
%\multicolumn{4}{||c||c||}{p=4 & p=5} \\
 \hline
 $N_E$ & $N_G$ & $N_C$ & $F$ & $N_G$ & $N_C$ & $F$ & $N_G$ & $N_C$ & $F$ \\ [0.5ex] 
\hline\hline
 2 & 6 & 0 & 0 & 10 & 0 & 0 &15 &0 &0\\ 
\hline
3 & 8 & 0 & 0 & 20 & 0 & 0 &40 &0 &0\\
\hline
4 & 27 & 6 & 0.22 & 85 & 0 & 0 &315 &0 &0\\
\hline
5 & 48 & 24 & 0.5 &  224 & 24 & 0.11 &744 &0 &0\\
\hline
6 & 112 & 64 &0.57 & 660 & 180 & 0.27 &2 770 &120 &0.04\\
\hline
7 & 192 & 144 & 0.75 & 1 640 & 720 & 0.44 &9 120 &1 440 & 0.16\\
\hline
8 & 378 & 291 & 0.77 & 4 095 & 2 285 & 0.56 &29 100 &8 370 &0.29\\
\hline
9 & 624 & 536  & 0.86 & 9 360 & 6 260  & 0.67 &86 600 &36 120 &0.42\\
\hline
10 & 1 092 & 954 & 0.87 & 20 910 & 15 470 & 0.74 &247 176 & 130 566 &0.53\\
\hline
11 & 1 728 & 1 584 & 0.92 & 44 220  & 35 520 & 0.80 &671 160 &417 960 &0.62\\
\hline 
12 & 2 802 & 2 593 & 0.93 & 90 945 & 76 825 & 0.84 &1 752 230 &1 223 520 &0.70\\
\hline
13 & 4 248 & 4 032 & 0.95 & 179 820 & 158 340  & 0.88 &4 396 200 &3 338 760 &0.76\\
\hline
14 & 6 516 & 6 216 & 0.95  & 346 320  &313 380 & 0.90 &10 655 670 &8 604 660 &0.81\\ 
\hline
15 & 9 528 & 9 216 & 0.97 & 646 860 & 598 680 & 0.93 & 24 983 264 &21 132 744 &0.85\\
\hline
\end{tabular}}
\caption{A table showing how many interaction graphs $N_G$ with $p$ vertices can be constructed using $N_E$ edges.
$N_C$ of the graphs are connected.
The fraction $F={N_C\over N_G}$ tells us the probability that a graph selected at random is connected.
The first three columns have $p=4$, the middle three columns have $p=5$ and the last three columns have $p=6$.}
\label{table:1}
\end{table}
The algorithm we use to generate interaction graphs is as follows:
\begin{itemize}
\item[1.] Partition the total number of edges $E$ in the graph into a sum of path lengths in all possible ways.
The Gauss constraint forces paths to have a length of at least 2.
For example, a graph with $E=4$ edges can be realized as two paths of length 2 or one path of length 4.
We assume that the interaction graph has a total of $p$ vertices and that these vertices are labeled as $1,2,...,p$.
\item[2.] Each path can be labeled with an ordered sequence of integers, which records the order in which the different
vertices are traversed as one travels on the path.
Each path visits any given vertex at most once.
Thus, the integers appearing in a given path label are distinct.
In addition, since the path is closed, cyclic shuffling of the integers in the path does not lead to a new path.
This makes it clear that the paths of length $L$ can be labeled by permutations that are a single cycle of length $L$.
We now need to sum over combinations of all possible paths consistent with the partition constructed in step 1.
\item[3.] The resulting list of interaction graphs will have some duplicates, since the decomposition of a given interaction graph 
into a collection of paths is not unique. 
The final step in the algorithm simple deletes the duplicate graphs.
\end{itemize}
For examples of the number of graphs obtained when using this algorithm, see Table \ref{table:1}.
It is noteworthy that the number of interaction graphs grows very rapidly.
For example, there are roughly 25 million interaction graphs with 15 edges and 6 vertices.
Such enormous numbers justify a statistical approach to the problem.

Before leaving this subsection, we will explain how to count the number of interaction graphs, using methods 
from information theory used to count Markov types\cite{MarkovTypes}.
This counting will enable us to understand the number of interaction graphs as $N\to\infty$.
Since the graph is a label for the operator, this will allow us to count the number of orthogonal\footnote{By orthogonal
operators, we mean operators which diagonalize the two point function. 
Thus they would be orthogonal in the Zamolodchikov norm of the conformal field theory.}
operators we have and thereby to verify that the growth is enough to explain the entropy of a black hole.
Introduce the matrix $E_{ij}$, $i,j=1,...,p$.
The off diagonal matrix elements $E_{ij}$ denote the number of edges running from vertex $i$ to vertex $j$.
Clearly $E_{ij}$ is not in general a symmetric matrix\footnote{The relation between $E_{ij}$ and the matrix $N_{ij}$
appearing in (\ref{GGham}) is $N_{ij}=E_{ij}+E_{ji}$.}.
The diagonal matrix elements vanish $E_{ii}=0$. 
Our task is to count the number of matrices obeying the equations
\bea
\sum_{i=1}^p E_{ij}=\sum_{i=1}^p E_{ji}\label{GC}
\eea
and
\bea
\sum_{i=1}^p\sum_{j=1}^p E_{ij}=N_E\label{totnum}
\eea
The equation (\ref{GC}) is the Gauss constraint and (\ref{totnum}) sets the number of edges in the graph.
The number of solutions to (\ref{GC}) and (\ref{totnum}) is the number of interaction graphs with $p$ vertices and $N_E$
edges, denoted $N_{p,N_E}$.

Let ${\cal E}$ be the set of all integer matrices obeying (\ref{GC}) and let ${\cal E}_{N_E}$ be the subset of matrices 
belonging to ${\cal E}$ that obeys (\ref{totnum}).
Given a pair of $p\times p$ matrices $E$ and $F$, we define
\bea
F^{*E}\equiv \prod_{\substack{i,j=1\\i\ne j}}^p F_{ij}^{E_{ij}}
\eea
We would like to evaluate the generating function
\bea
Z_{\rm Gauss}(F)=\sum_{E\in{\cal E}}F^{*E}=\sum_{n_E\ge 0}\sum_{E\in{\cal E}_{n_E}}F^{*E}
\eea
Evaluating this generating function at $F_{ij}=z$ and using the obvious fact
\bea
\sum_{E\in{\cal E}_{n_E}}F^{*E}=\sum_{E\in{\cal E}_{n_E}}z^{\sum_{i=1}^p\sum_{j=1}^pE_{ij}}
=\sum_{E\in{\cal E}_{n_E}}z^{n_E}=N_{p,N_E}z^{n_E}
\eea
we find
\bea
Z_{\rm Gauss}(F_{ij}=z)=\sum_{n_E\ge 0} N_{p,N_E}z^{N_E}
\eea

We will now give a useful integral representation for $Z_{\rm Gauss}(F)$ that uses nothing more than the residue theorem.
First, introduce the diagonal matrix
\bea
D_X=\left[
\begin{matrix}
x_1     &0             &\cdots &0          &0\\
0        &x_2          &\cdots &0          &0\\
\vdots &\vdots      &\ddots &\vdots   &\vdots\\
0        &0             &\cdots &x_{p-1} &0\\
0        &0             &\cdots &0           &x_p
\end{matrix}\right]
\eea
which we will use below.
Next, introduce the generating function
\bea
Z(F)=\sum_{E} F^{*E}
\eea
where the sum above is over all matrices $E_{ij}$ with zeros on the diagonal and non-negative integers off the diagonal.
There are two reasons for why it is useful to introduce this new generating function.
First, it is a simple task to evaluate the sum and obtain an explicit answer 
\bea
Z(F)=\prod_{\substack{i,j=1\\ i\ne j}}^p (1-F_{ij})^{-1}
\eea
Second, it is possible to express $Z_{\rm Gauss}(F)$ as a contour integral over $Z(F)$.
To see this, note that the term that is independent of $x_i$, $i=1,\cdots ,p$ in
\bea
Z(D_X^{-1}FD_X)&=&\sum_{E} \prod_{\substack{i,j=1\\ i\ne j}}^p F_{ij}^{E_{ij}}
\prod_{k=1}^p x_k^{\sum_{l=1,l\ne k}^p E_{kl}-\sum_{l=1,l\ne k}^p E_{lk}}\cr
&=&\prod_{\substack{i,j=1\\ i\ne j}}^p \left(1-z{x_i\over x_j}\right)^{-1}
\eea
is obviously $Z_{\rm Gauss}(F)$.
Thus we have
\bea
Z_{\rm Gauss}(F_{ij}=z)=\left({1\over 2\pi i}\right)^p
\oint {dx_1\over x_1}\cdots\oint {dx_p\over x_p}
\prod_{\substack{i,j=1\\ i\ne j}}^p \left(1-z{x_i\over x_j}\right)^{-1}\label{GGgen}
\eea
As an example, when $p=4$ we find
\bea
Z_{\rm Gauss}(F_{ij}=z)&=&
\frac{z^8-2 z^7+3 z^6+2 z^5-2 z^4+2 z^3+3 z^2-2 z+1}{(1-z)^9 (z+1)^5\left(z^2+1\right)\left(z^2+z+1\right)^2}\cr\cr
&=&1+6 z^2+8 z^3+27 z^4+48 z^5+112 z^6+192 z^7+378 z^8+624 z^9+1092 z^{10}\cr
&&+1728 z^{11}+2802 z^{12}+4248 z^{13}+6516 z^{14}+9528 z^{15}+O(z^{16})
\eea
which nicely confirms our numerical results in Table \ref{table:1}.
For $p=5$ we have
\bea
Z_{\rm Gauss}(F_{ij}=z)={n(z)\over (1 - z)^{16}(1 + z)^8 (1 + z^2)^2 (1 + z + z^2)^4
(1 + z + z^2 +z^3 + z^4)}\label{pis5}
\eea
where
\bea
n(z)&=&z^{20} - 3 z^{19} + 7 z^{18} + 3 z^{17} + 2 z^{16} + 17 z^{15} + 
 35 z^{14} + 29 z^{13} + 45 z^{12} + 50 z^{11} + 72 z^{10}\cr
&& + 50 z^9 + 
 45 z^8 + 29 z^7 + 35 z^6 + 17 z^5 + 2 z^4 + 3 z^3 + 7 z^2 - 3 z + 1
\eea
Expanding (\ref{pis5}) we again confirm the results in Table \ref{table:1}.

Starting from (\ref{GGgen}) we can now explore the growth of the number of interaction graphs as we take $p\to\infty$.
Setting $x_i=e^{i\theta_i}$ we have
\bea
Z_{\rm Gauss}(F_{ij}=z)=\left({1\over 2\pi}\right)^p
\int_{-\pi}^\pi d\theta_1 \cdots\int_{-\pi}^\pi d\theta_p
\prod_{\substack{i,j=1\\ i\ne j}}^p \left(1-ze^{i(\theta_i-\theta_j)}\right)^{-1}
\eea
The integrand is invariant under the simultaneous shift $\theta_i\to\theta_i-a$, $i=1,2,\cdots,p$.
Using this symmetry to carry out the integral over $\theta_1$ we obtain
\bea
Z_{\rm Gauss}(F_{ij}=z)&=&\left({1\over 2\pi}\right)^{p-1}
\int_{-\pi}^\pi d\theta_2 \cdots\int_{-\pi}^\pi d\theta_p
\prod_{i=2}^p \left(1-ze^{i\theta_i}\right)^{-1}\left(1-ze^{-i\theta_i}\right)^{-1}\cr
&\times& \prod_{\substack{i,j=2\\ i\ne j}}^p \left(1-ze^{i(\theta_i-\theta_j)}\right)^{-1}
\eea
In terms of the function
\bea
L(z,\theta_2,\cdots,\theta_p)=\sum_{i=2}^p \log\left[
\left(1-ze^{i\theta_i}\right)\left(1-ze^{-i\theta_i}\right)\right]
+\sum_{\substack{i,j=2\\ i\ne j}}^p \log\left[1-ze^{i(\theta_i-\theta_j)}\right]
\eea
we can write the number of interaction graphs as
\bea
N_{p,N_E}={1\over i\, (2\pi)^p}\oint {dz\over z^{1+N_E}}\int_{-\pi}^\pi d\theta_2 \cdots\int_{-\pi}^\pi d\theta_p
e^{-L(z,\theta_2,\cdots,\theta_p)}
\eea
To determine the asymptotic behavior of this integral we will use a saddle point evaluation as usual.
Using the equivalent form
\bea
L(z,\theta_2,\cdots,\theta_p)=\sum_{i=2}^p \log\left[1-2z \cos \theta_i+z^2\right]
+{1\over 2}\sum_{\substack{i,j=2\\ i\ne j}}^p \log\left[1-2z\cos (\theta_i-\theta_j)+z^2\right]
\eea
it is simple to verify that $L(z,\theta_2,\cdots,\theta_p)$ assumes its minimum value at $\theta_2=\cdots=\theta_p=0$.
An equally simple computation shows that, at this minimum, $L(z,\theta_2,\cdots,\theta_p)+(N_E+1)\log z$ is minimized at
\bea
z={N_E+1\over p^2-p+N_E+1}
\eea
Setting $p=\epsilon N$ and $N_E=\epsilon^2 N^2$ and working to leading order in the saddle point approximation, we find
at large $N$ that
\bea
N_{p,N_E}\sim e^{2\epsilon^2 N^2\log (2)} 
\eea
Assuming that the interaction graphs do indeed label microstates of a black hole, this is the correct growth to reproduce the
expected black hole entropy.

\subsection{Characterizing interaction graphs}

Given this algorithm we can now easily generate collections of graphs, and then use these to numerically characterize 
the properties of interaction graphs.
We would like to employ the notion of {\it typicality}.
Something is typical if it happens in the vast majority of cases: the typical lottery ticket loses, after 1000 coin flips
we typically find the ratio of the number of heads to the number of tails is close to 1 and so on.
We would like to characterize the typical interaction graph.

Our goal now is to make the above intuitive notions mathematically precise.
For useful background see \cite{typicality}. 
What does it mean for an interaction graph to be typical?
Consider an element $x$ of a set $S$, $x\in S$.
Typicality is a relational property of $x$, which $x$ possesses with respect to $S$. 
Typicality refers to an attribute $P$ and a (probability) measure for this attribute $\mu_P$.
For our discussion, $S$ is the set of all interaction graphs, with a given number of vertices $p$ and edges $E$, denoted 
$S_{p,E}$.
As discussed in the previous section, we can consider $p\sim\epsilon N$.
Thus, at large $N$ we know that $p$ is enormous and the number of interaction graphs explodes.
We will also assume that we are in the ``dense graph'' regime specified by allowing the total number of edges to scale
as $E\sim p^2\sim \epsilon^2 N^2$.
Thus, we are interested in characterizing the typical graph in the set $S_{p,p^2}$ of interaction graphs.

We will define the measure $\mu_P$ simply by counting.
This assumes that every graph is equally likely.
In this case the probability $\mu_P$ that a given graph has property $P$ is simply given by counting the number of graphs 
with property $P$ and then dividing by the total number of graphs.
When $\mu_P$ tends towards 1, $P$ becomes a property of a typical graph.
In what follows we are interested in determining some of the properties of a typical graph in $S_{p,p^2}$. 

One interesting attribute $P$ is whether or not the graph is connected. 
For a Hamiltonian defined using a disconnected graph, the bosons hopping on the graph are confined to a given 
connected component.
A state that is not initially entangled can never build up entanglement between Hilbert spaces defined on vertices of different
disconnected components of the graph.
Its only on a connected graph that an initial state that is not entangled can evolve into a maximally entangled state,
entangling all of the Hilbert spaces defined at the different vertices.
The trend shown in Table \ref{table:1} is exactly what one expects: for a fixed number of vertices, as the number of 
edges increases the probability that the graph is connected (denoted by $F$ in Table \ref{table:1}) increases. 
Our numerical results imply that just as $E$ approaches $p^2$, this probability of being connected approaches 
1\footnote{For $p=4$ and $E=20$ we find that a graph is connected with probability 0.99.}.
With this numerical evidence, we assume in what follows that the typical graph in $S_{p,p^2}$ is connected.

\begin{figure}[h]
     \centering
     \includegraphics[width=0.95\textwidth]{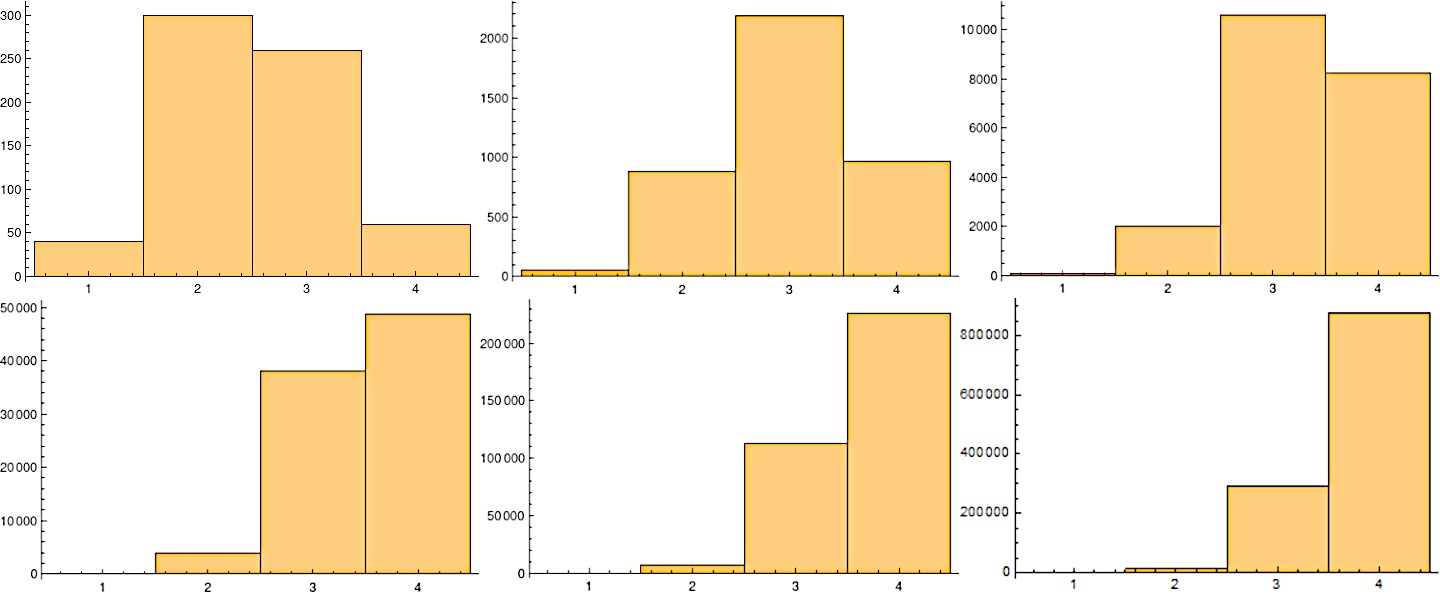}
     \caption{The above plots show histograms of the maximum vertex degree $D_V$, for Gauss graphs with $5$ vertices and
                  $E=6,8,10,12,14$ and 16 edges.}
     \label{fig:MaxVert}
\end{figure}

As discussed in the introduction, when deriving the Lieb-Robinson bound for dynamics on a graph an important parameter
which enters the bound is the maximum degree $D_V$ of any vertex in the graph\footnote{Our graphs can have multiple 
edges between a given pair of vertices. $D_V$ counts how many other vertices $V$ is connected to and not the number of
edges with an endpoint on $V$}.
Thus, a second interesting attribute $P$ for the questions we consider is the maximum degree $D_V$.
In Figure \ref{fig:MaxVert} we have given histograms for the different values of $D_V$ on the sets $S_{5,E}$.
In this case, the largest value $D_V$ can attain is 4, when a given vertex connects to all of the remaining vertices.
The first histogram has $E=6$ edges.
There are significant fractions of graphs with all possible allowed values $D_V=1,2,3,4$.
As $E$ increases a definite pattern emerges: $D_V=4$ becomes the most probable value for $E\ge 12$ edges.
The largest value shown is $E=16$ edges. 
It is clear that by the time we reach $E=p^2=25$ edges, the overwhelming majority of graphs will have $D_V=4=p-1$.
Based on this numerical evidence, we assume in what follows that the typical graph in $S_{p,p^2}$ has $D_V=p-1$.
Since we work at large $p$ we simplify this to $D_V=p$.
\vspace{0.75cm}

It is interesting to ask how this typical value of $D_V$ is reached. It maybe that most graphs have a single vertex with a
large value for $D_V$ and the remaining vertices have much smaller values for their degree.
In this case, since there are $p$ vertices, we will find that the average vertex degree stays close to 1.
The opposite extreme is that the degree of all vertices is increasing roughly equally, so that most graphs in $S_{p,p^2}$
have an average vertex degree which is close to the maximum value of $p-1$.
Numerically we find that the average vertex degree is an increasing function with the number of edges $E$ (see
Figure \ref{fig:AvVert}) and that when $E\sim p^2$ we find an average value close to the maximum allowed value.
Of course we can not probe large values of $p$ (already $p=6$ requires very long run times), but this conclusion makes 
sense: nothing has introduced an asymmetry between the $p$ vertices, so we would expect the degree of each vertex to
be roughly equal.
Thus, from now on we assume that most vertices in a typical graph in $S_{p,p^2}$ have the maximum degree.
\begin{figure}[h]
     \centering
     \includegraphics[width=0.5\textwidth]{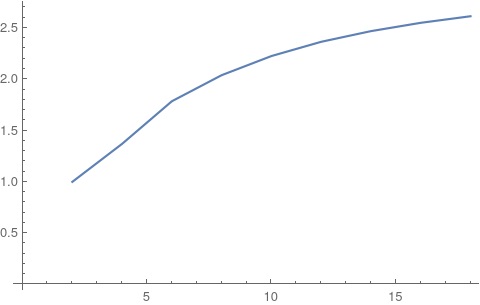}
     \caption{The above plots show the average vertex degree for graphs with $p=4$ vertices versus the number of edges $E$.
                  For $E=16$ the average vertex degree is 2.55. This indicates that most vertices that have been averaged over must
                  assume the maximum value of $p-1=3$.}
     \label{fig:AvVert}
\end{figure}

The conclusions we have reached in this section regarding the typical interaction graph have a number of interesting implications.
Recall that each vertex in the graph is a giant graviton brane and each edge is an open string excitation of the brane.
By characterizing the typical graph we are learning about the typical excited state of this $p$ giant graviton system.
The typical state of a system of $p$ giant graviton branes, excited by stretching $p^2$ open strings between the branes,
has roughly the same number of open strings endpoints glued to each brane.
In terms of the Gauss graph operators, the differences between the row lengths of $R$ and those of $r$ are roughly constant,
equal to $p$.
This is good news: if one simply piled all the excitations into a small number of rows of $R$\footnote{This would correspond to
piling many edges onto one vertex of the interaction graph.} one might imagine a situation
in which Young diagram $R$ satisfies the distant corners approximation, but the approximation breaks down for $r$.
This might invalidate the derivation given in \cite{deMelloKoch:2020agz} which assumes that the corners of both $R$ 
and $r$ are distant.
Fortunately this does not happen for the typical graph which justifies the distant corners approximation.

Many proposed quantum mechanical models of black holes include highly non-local interactions.
A good example is the SYK model \cite{Kit,SaYe,Maldacena:2016hyu} which is a lattice model with all-to-all interactions.
Since each vertex of a typical graph of $S_{p,p^2}$ has a mean vertex degree which is close to the maximum value,
we find that the generic Hamiltonian defined by the mixing problem for heavy operators, has all-to-all interactions.

\section{Scrambling on typical graphs}\label{scramble}

In this section we would like to explore how quickly entanglement is generated by the typical graph Hamiltonian.
We restrict ourselves to the subspace of Fock space with a definite number $N_b$ of bosons.
Towards this end, in the next section we will formulate a Lieb-Robinson bound for the typical graph Hamiltonian.
The bound limits the growth of commutators $[O_i(0),O_j(t)]$ where $O_i\in{\cal F}_i$ so that we are bounding the
growth of operators \cite{Roberts:2014isa}.
The growth of operators is a reliable probe of scrambling\cite{Lensky:2018hwa,Lashkari:2011yi,Hosur:2015ylk,Sahu:2020xdn}.
To obtain our bounds, we use arguments of \cite{HarmLR}, used to derive Lieb-Robinson bounds for general 
harmonic systems on general lattices.
Information that has been scrambled is stored in the complicated correlations between many different subsystems.
Consequently, scrambling is intimately related to the generation of entanglement.
With this motivation, we consider in Section \ref{entgen} a toy model for our system, simple enough that we can 
compute the Von Neumann entropy as a function of time, using the reduced density matrix of a given Fock space 
${\cal F}_i$ and starting from an initially unentangled state.
Although this explicitly shows the generation of entanglement in the system, it also poses a puzzle: the recurrence
time associated with the typical Hamiltonian is much smaller than expected.

\subsection{Lieb-Robinson bound for typical graph dynamics}\label{LRBound}

Trade the oscillator operators $b_i,b_i^\dagger$ for a pair of Hermittian operators, $\alpha_i$ and $\beta_i$ given by
\bea
\alpha_i={b_i+b_i^\dagger\over\sqrt{2}}\qquad\qquad\qquad \beta_i={b_i-b_i^\dagger\over i\sqrt{2}}
\eea
This is a complete set of operators in the sense that if for any operator $O$ we have
\bea
[\alpha_i ,O]=0\qquad [\beta_i,O]=0
\eea
then $O$ is a multiple of the identity operator.
Rewriting the Hamiltonian in terms of the $\alpha_i,\beta_i$ operators, we obtain the following result
\bea
H=H_0+{1\over 2}\sum_{i,j=1}^p \alpha_i M_{ij}\alpha_j+{1\over 2}\sum_{i,j=1}^p \beta_i M_{ij}\beta_j
\eea
where $H_0$ is an additive constant equal to
\bea
H_0={2g_{YM}^2\over (4\pi)^2}\sum_{i,j=1}^p(\sqrt{r_i}-\sqrt{r_j})^2 N_{ij}
        -{g_{YM}^2\over 4\pi^2}\sum_{i=1}^p {r_i k_i\over l_i}
\eea
and the matrix $M_{ij}$ is given by
\bea
M_{ij}={2g_{YM}^2\over (4\pi)^2}{r_i k_i\over l_i}\delta_{ij}
-{2g_{YM}^2\over (4\pi)^2}\sqrt{r_i r_j\over l_il_j}N_{ij}
\eea
Recall that $r_i$, $l_i$ and $k_i$  were introduced in Section \ref{dynamics}.
A simple computation shows that
\bea
\alpha_i (t)&=&e^{iHt}\alpha_i e^{-iHt}\cr
&=&\left[\cos (Mt)\right]_{kj}\alpha_j +\left[\sin (Mt)\right]_{kj}\beta_j
\label{asol}
\eea
\bea
\beta_i (t)&=&e^{iHt}\beta_i e^{-iHt}\cr
&=&\left[\cos (Mt)\right]_{kj}\beta_j -\left[\sin (Mt)\right]_{kj}\alpha_j
\label{bsol}
\eea
Using these results we immediately obtain the following commutators
\bea
i\left[ \alpha_k (t),\beta_j\right]&=&-\,\cos (Mt)_{kj}\qquad
i\left[ \alpha_k (t),\alpha_j\right]=\sin (Mt)_{kj}\cr\cr
i\left[ \beta_k (t),\alpha_j\right]&=&\cos (Mt)_{kj}\qquad
i\left[ \beta_k (t),\beta_j\right]=\sin (Mt)_{kj}\label{bbcom}
\eea
To proceed we would like to estimate the size of terms of the form $(M^n)_{kj}$.
Recall that $p=\epsilon N$.
From our analysis of the typical graph, we know that all vertex degrees are close to the maximal value of $p$, which
implies that matrix elements $N_{ij}$ are typically non-zero and order $1$.
Consequently, the size of the off diagonal elements of $M_{ij}$ are
\bea
-{2g_{YM}^2\over (4\pi)^2}\sqrt{r_i r_j\over l_il_j}N_{ij}\sim
-{2g_{YM}^2\over (4\pi)^2}\sqrt{r_i r_j\over l_il_j}=
-{2\epsilon \lambda\over p (4\pi)^2}\sqrt{r_i r_j\over l_il_j}\equiv {\epsilon\lambda c_{ij}\over p}
\eea
where
\bea
c_{ij}=-{2\epsilon \lambda\over (4\pi)^2}\sqrt{r_i r_j\over l_il_j}
\eea
is a small number, independent of $N$.
For the diagonal elements of $M_{ij}$, we use the fact that for the typical graph we have $k_i=$ the degree of the
$i$th vertex $=p$ independent of $i$ and hence these matrix elements are of size
\bea
{2g_{YM}^2\over (4\pi)^2}{r_i k_i\over l_i}
={2\epsilon \lambda \over (4\pi)^2}{r_i\over l_i}\equiv \epsilon\lambda c_{i}
\eea
which is independent of $N$.
Using these results, we can bound the size of $|(M^n)_{ij}|$, for $i\ne j$.
Choose the constant $c>0$ to be larger than $|c_{ij}|$ for all $i,j$ and larger than $|c_i|$ for all $i$.
We will illustrate the computation with two examples and then state the general rule.
For $n=1$ we are talking about an off diagonal element so that
\bea
M_{ij}={\lambda\epsilon c_{ij}\over p}\qquad\Rightarrow\qquad |M_{ij}|< {c\lambda\epsilon\over p}
\eea
For $n=2$ we have a product of two matrices. There is a single index summed.
Thus, we have $p-2$ terms which are off diagonal and two terms that are the product of an off diagonal element with a 
diagonal element, so that
\bea
|(M^2)_{ij}|&=&\left|\sum_{\substack{k=1\\k\ne i,j}}^p
{c_{ik}c_{kj}\epsilon^2\lambda^2\over p^2}+{c_{ij}\epsilon^2\lambda^2\over p}(c_{ii}+c_{jj})\right|\cr
&<& \sum_{\substack{k=1\\k\ne i,j}}^p
{|c_{ik}|\, |c_{kj|}\,\epsilon^2\lambda^2\over p^2}+{|c_{ij}|\,(|c_{ii}|+|c_{jj}|)\,\epsilon^2\lambda^2\over p}\cr
&<&(p-2){c^2\epsilon^2\lambda^2\over p^2}+2{c^2\epsilon^2\lambda^2\over p}=3 {c^2\epsilon^2\lambda^2\over p}
+O\left({1\over p^2}\right)
\eea
We drop the $p^{-2}$ term.
Proceeding in this way its easy to see that
\bea
|(M^n)_{ij}| < (2n+1){c^n\epsilon^n\lambda^n\over p}
\eea
Consequently, for example, we can estimate
\bea
|i[\alpha_k(t),\beta_j]|=
|\cos (Mt)_{kj}|&<&\sum_{n=0}^\infty {t^{2n}\over (2n)!}|(M^{2n})_{kj}|\cr
&<&\sum_{n=0}^\infty {t^{2n}\epsilon^{2n}\lambda^{2n}\over (2n)!}{c^{2n}\over p} (2n+1)\cr
&=&{\cosh (c\epsilon\lambda t)+c\epsilon\lambda t\sinh (c\epsilon\lambda t)\over p}
\eea
The right hand side becomes of order $1$ when $e^{c\epsilon\lambda t}\sim p$ i.e. when 
$t\sim {\log p\over \epsilon\lambda c}$.
At this time scale, the bounds for all of the commutators in (\ref{bbcom}) are order $1$.

Define the amount of time $t_{\rm sig}$ as the smallest time needed to signal from site $j$ to site $k$.
This implies that for suitable operators $O_j$ acting on ${\cal F}_j$ and $O_k$ acting on ${\cal F}_k$ we have
\bea
\langle\psi(0)|[O_j(0),O_k(t_{\rm sig})]|\psi(0)\rangle >\delta
\eea
with $\delta$ some $O(1)$ number.
The Lieb-Robinson bound then forces 
\bea
t_{\rm sig}>  {\log p\over \epsilon\lambda c}
\eea
This logarithmic scaling of signaling implies a logarithmic scaling of the scrambling time.
Thus, our Hamiltonian does not scramble in a time less than $\sim \log p$ consistent with 
the fast scrambling conjecture \cite{Sekino:2008he}.

\subsection{Entanglement Generation}\label{entgen}

Recall that the matrix $M_{ij}$ for the typical interaction graph is given by
\bea
M_{ij}=v_i v_j\left(\delta_{ij}-{1\over p}\right)\qquad\qquad 
v_i=\sqrt{{2\lambda\epsilon\over (4\pi)^2}}\sqrt{{r_i\over l_i}}
\eea
where $\epsilon,\lambda$ are both fixed and much smaller than 1 as we take $N\to\infty$.
The ratios ${r_i\over l_i}$ are larger than 1 and fixed as we take $N\to\infty$.
We will now make a simplifying assumption, that will yield a problem that is simple enough to solve.
We assume that the $v_i$'s are so similar that we can simply set them to be equal to $v$.
There are examples for which this is indeed an accurate assumption, but this is besides the point.
We make the assumption because it leads to a simple model that nevertheless captures the scaling with
$p$ of matrix elements of the Hamiltonian and it also captures the all-to-all interactions property of the 
typical Hamiltonian.
In this case $M$ has the following form
\bea
M=v^2(1+{1\over p}){\bf 1}+v^2 K
=v^2 (1+{1\over p}){\bf 1}+v^2\left[
\begin{array}{cccccc}
-{1\over p} &-{1\over p} &-{1\over p} &\hdots &-{1\over p} &-{1\over p}\cr
-{1\over p} &-{1\over p} &-{1\over p} &\hdots &-{1\over p} &-{1\over p}\cr
-{1\over p} &-{1\over p} &-{1\over p} &\hdots &-{1\over p} &-{1\over p}\cr
\vdots &\vdots &\vdots &\ddots &\vdots &\vdots\cr
-{1\over p} &-{1\over p} &-{1\over p} &\hdots &-{1\over p} &-{1\over p}\cr
-{1\over p} &-{1\over p} &-{1\over p} &\hdots &-{1\over p} &-{1\over p}
\end{array}
\right]
\eea
where ${\bf 1}$ is the $p\times p$ identity matrix.
Starting from the initial state
\bea
|\psi(0)\rangle =b_i^\dagger |0\rangle
\eea
it is straight forward to find the probability that we have a particle on site $i$ at time $t$
\bea
|\langle 0|b_i|\psi(t)\rangle|^2=\frac{p^2+2 (p-1) \cos (p v^2 t)-2 p+2}{p^2}\equiv p_i(t)
\eea
The reduced density matrix obtained by tracing over all Fock spaces ${\cal F}_j$, $j\ne i$ acts on the two dimensional 
subspace of ${\cal F}_i$ with basis $\{|0\rangle,b_i^\dagger |0\rangle\}$. 
The reduced density matrix is given by
\bea
\rho_i(t)=\left[
\begin{array}{cc}
1-p_i(t) &0\cr
0          &p_i(t)
\end{array}
\right]
\eea
and the corresponding Von Neumann entropy is
\bea
S_i&=&-\Tr\left(\rho_i\log\rho_i\right)\cr\cr
&=&-\frac{2 (p-1) \cos (v^2 t)+p^2-2p+2}{p^2} \log \left(\frac{2 (p-1) \cos (v^2 t)+p^2-2p+2}{p^2}\right)\cr
&&+\frac{2 (p-1) (1-\cos (v^2 t))}{p^2} \log \left(\frac{2 (p-1) (1-\cos (v^2 t))}{p^2}\right)
\eea

\vspace{0.5cm}

\begin{figure}[h]
     \centering
     \includegraphics[width=0.5\textwidth]{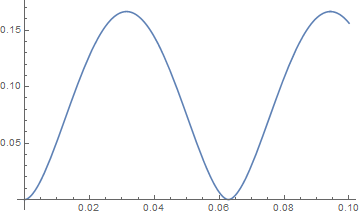}
     \caption{The above plot shows the Von Neumann entropy versus $v^2 t$, for $p=100$.}
     \label{fig:AvVert}
\end{figure}
The above plot shows that entanglement is generated and it exhibits the recurrence time of $T_r=2\pi/v^2$.
At time $T_r$ the entanglement entropy returns to zero and the curve repeats.
This recurrence time is much smaller than expected and it casts a doubt on the model.
Typically the recurrence is doubly exponential in the size of the system \cite{Page:1994dx,venuti,Susskind:2018pmk}.
This suggests a recurrence time of $\sim \exp\exp p$.
In the next section we will consider equilibration for dynamics described by the typical Hamiltonian, which will lead
to an explanation for this tiny recurrence time.

\section{Equilibration during intervals}\label{equal}

The classical idea of equilibration involves evolution towards thermal equilibrium.
A closed finite dimensional quantum system evolving unitarily has recurrent and time reversal invariant dynamics.
It can never come to equilibrium in the classical sense, so we must relax the above classical notion.
At quantum level, equilibration will mean that a quantity, initialised at a non-equilibrium value, evolves towards the 
equilibrium value and then stays close to it for an extended time\footnote{This requires a measure that quantifies how 
close the value of an obervable is to its ``equilibrium value”.}.
This leads to two natural notions of equilibration (see \cite{equilibration}):
\begin{itemize}  
\item[1.]{\bf Equilibration on average:} A time dependent observable equilibrates on average if its value is for most times 
during the evolution close to some equilibrium value.
\item[2.]{\bf Equilibration during intervals:} A time dependent property equilibrates during a (time) interval if its 
value is close to some equilibrium value for all times in that interval.
\end{itemize}
Results establishing equilibration during intervals imply bounds on the time it takes to equilibrate.
These time scales are of central interest to us, so we use the second notion above.
The conditions under which equilibration during intervals of quadratic bosonic Hamiltonians\footnote{Hamiltonians 
that are quadratic polynomials in the bosonic creation and annihilation operators.} can be guaranteed has been studied 
in \cite{Cramer:2008zz,quadequil}. 
Since our system is a quadratic bosonic Hamiltonian, these results are immediately applicable.

We will study a subsystem, given for simplicity by a single site.
The rest of the system behaves like a heat bath allowing the subsystem to reach a state that maximizes its entropy. 
Our strategy is to start the entire lattice system in an initial non-equilibrium state and then to demonstrate that
the state of a single site evolves to the equilibrium state, that is, the state that maximizes the entropy.

To carry out the computation, we need to know something about the state that maximizes the entropy.
The state (density operator $\rho$) that maximizes the entropy 
\bea
H(\rho)=-\Tr (\rho\log\rho )
\eea
for given mean and second moments\footnote{Define $b(\alpha)=\sum_{i=1}^l (\alpha_i b_i^\dagger -\alpha_i^* b_i)$. 
We call $\langle b(\alpha)\rangle$ the mean and $\langle b(\alpha)b(\alpha')\rangle$ the correlation matrix or the 
second moment.}, is a Gaussian state \cite{HSH}.
Instead of discussing the density operator itself, it is useful to study the characteristic function $\chi(\beta)$.
The characteristic function contains all the information necessary to reconstruct the density matrix so that it is an
alternative description of the system.
The characteristic function is given by the expectation value of the Weyl operators, defined by
\bea
D(\beta)=e^{\sum_{i=1}^p \beta_i b^\dagger_i-\beta_i^* b_i}
\eea
A density operator $\rho$ is called Gaussian if its quantum characteristic function has the form
\bea
\Tr (\rho\, V(z)) = e^{i m\cdot \beta -{1\over 2}\beta^* \cdot \alpha\cdot \beta}
\eea
with $m_i,\alpha_{ij}$ constants independent of $\beta_i,\beta_i^*$.
To see why Gaussian states maximize the entropy, recall that given any density matrix $\rho$, there is a Gaussian density 
matrix $\tilde\rho$ with the same mean and second moments\cite{holevo}.
Consider the quantity
\bea
H(\tilde\rho)-H(\rho)=\Tr \left(\rho(\log\rho-\log\tilde\rho)\right)+\Tr\left( (\rho-\tilde\rho)\log\tilde\rho\right)
\eea
The first term on the right hand side is the relative entropy, which is non-negative\cite{relEn}.
The second term on the right hand side vanishes because (i) $\log\tilde\rho$ is a quadratic polynomial in $b_i,b_i^\dagger$
and (ii) $\rho$ and $\tilde\rho$ have the same first and second moments.
This proves that
\bea
H(\tilde\rho)-H(\rho)\ge 0
\eea
which proves the statement.

We now study the reduced density matrix for a singe site.
The environment $E$ is all sites except for site $i$.
The reduced density matrix
\bea
\rho_i = \Tr_E (|\phi\rangle\langle\phi|)
\eea
is an operator acting in the Hilbert space associated to the $i$th site.
Put $m$ bosons on each site so that the initial state is
\bea
|\phi\rangle = |m\rangle^{\otimes p}\qquad |m\rangle^{\otimes p} =\prod_{i=1}^p {(b_i^\dagger)^m\over\sqrt{m!}}
|0\rangle_i
\eea
Notice that this initial state is not entangled and is nothing like the maximally entangled equilibrium state.
We will evaluate the characteristic function
\bea
\chi_i (\alpha,t)
=\Tr \left(\rho_i(t)\, e^{\alpha b_i^\dagger - \alpha^* b_i}\right)
=\langle\phi |e^{\alpha b_i^\dagger(t) - \alpha^* b_i(t)}|\phi\rangle
\eea
From (\ref{asol}) and (\ref{bsol}) we have
\bea
b_i(t)=(e^{-itM})_{ij} b_j\equiv U_{ij}b_j\qquad\qquad
b_i^\dagger (t)=(e^{itM})_{ji} b_j^\dagger =U^*_{ji}b_j^\dagger
\eea
$M$ is a symmetric matrix so $U$ is also symmetric.
Using the initial state
\bea
\chi_i(t)=\langle\phi|e^{\sum_{j=1}^p(\alpha U_{ij}b_j-\alpha^* U_{ij}^* b_j^\dagger)}|\phi\rangle
=\prod_{j=1}^p {}_j\langle m|e^{\alpha U_{ij}b_j-\alpha^* U_{ij}^* b_j^\dagger}|m\rangle_j
\eea
we can evaluate each factor in this product. 
First, using the Baker-Campbell-Haussdorf formula \cite{BCH} it is simple to verify that (no sum on $j$)
\bea
e^{\alpha U_{ij}b_j-\alpha^* U_{ij}^* b_j^\dagger}=
e^{\alpha U_{ij}b_j}e^{-\alpha^* U_{ij}^* b_j^\dagger}e^{-{|\alpha|^2\over 2}|U_{ij}|^2}
\eea
Using this identity we easily find
\bea
{}_j\langle m|e^{\alpha U_{ij}b_j-\alpha^* U_{ij}^* b_j^\dagger}|m\rangle_j
&=&{}_j\langle m|e^{\alpha U_{ij}b_j}e^{-\alpha^* U_{ij}^* b_j^\dagger}|m\rangle_j
e^{-{|\alpha|^2\over 2}|U_{ij}|^2}\cr
&=&\sum_{k=0}^m\sum_{l=0}^m {(\alpha U_{ij})^k(-\alpha^* U_{ij}^*)^l \over k! l!}
{m!\over \sqrt{(m-k)! (m-l)!}}{}_j\langle m-k|m-l\rangle_j e^{-{|\alpha|^2\over 2}|U_{ij}|^2}\cr
&=&\sum_{k=0}^m {(-|\alpha|^2 |U_{ij}|^2)^k \over k! k!}
{m!\over (m-k)!} e^{-{|\alpha|^2\over 2}|U_{ij}|^2}\cr
&=&L_m(|\alpha|^2 |U_{ij}|^2)e^{-{|\alpha|^2\over 2}|U_{ij}|^2}
\eea
where $L_m(\cdot)$ is a Laguerre polynomial.
Thus, we find
\bea
\chi_i(t)=\prod_{j=1}^p L_m(|\alpha|^2 |U_{ij}|^2)e^{-{|\alpha|^2\over 2}|U_{ij}|^2}
\eea

We want to prove that, at late times, the characteristic function becomes a Gaussian, i.e. that at late times we have
\bea
\chi_i(t)=e^{-c\alpha^2}
\eea
where $c$ is a constant (independent of $\alpha$).
Consider
\bea
\log\chi_i(t) =-\sum_{j=1}^p {|\alpha^2|\over 2}|U_{ij}|^2+\sum_{j=1}^p\log L_m(|\alpha|^2 |U_{ij}|^2)
\eea
Expand the log
\bea
\sum_{j=1}^p\log L_m(|\alpha|^2 |U_{ij}|^2)
=\sum_{j=1}^p\left(1-L_m(|\alpha|^2 |U_{ij}|^2)\right)+\sum_{j=1}^p\sum_{k=2}^\infty
{\left(1-L_m(|\alpha|^2 |U_{ij}|^2)\right)^k\over k}
\eea
and use the expansion of the Laguerre polynomials
\bea
L_m(x)=\sum_{n=0}^m {m!\over (m-n)! n! n!}(-x)^n =1-mx+O(x^2)
\eea
If we can argue that $x=|\alpha|^2 |U_{ij}|^2$ is small for late times $t$, then we can set
\bea
\sum_{j=1}^p\log L_m(|\alpha|^2 |U_{ij}|^2)
=\sum_{j=1}^p m |\alpha|^2 |U_{ij}|^2
\eea
and consequently
\bea
\log\chi_i(t)
=-\sum_{j=1}^p {|\alpha^2|\over 2}|U_{ij}|^2-\sum_{j=1}^p m\, |\alpha|^2 |U_{ij}|^2
\eea
which would establish the result.
Determining how rapidly $|\alpha|^2 |U_{ij}|^2$ approaches zero will tell us how quickly the system equilibrates.

Why should $|\alpha|^2 |U_{ij}|^2$ get small for large $t$? 
The matrix $M_{ij}$ is real and symmetric, so that it can be diagonalized.
Denote the eigenvectors and eigenvalues of $M_{ij}$, labeled by  $k=1,\cdots p$, as $(\eta_k)_i$ and $\lambda_k$ respectively.
In terms of these eigenvectors and eigenvalues we have
\bea
U_{ij}=\sum_{k=1}^p (\eta_k)_i\,e^{-i\lambda_k t}\, (\eta^*_k)_j
\eea
so that
\bea
|U_{ij}|^2 =\sum_{k=1}^p\sum_{l=1}^p \,\, (\eta_k)_i\,e^{-i\lambda_k t}\, (\eta^*_k)_j\,\,
(\eta^*_l)_i\,e^{i\lambda_l t}\, (\eta_l)_j\label{togetsmall}
\eea
There are two sums above which become infinite sums at large $N$.
If the eigenvalues are distinct, we are adding terms with different rapidly oscillating phases for large $t$, so that
there will be many cancellations and we expect the sum is small.
To formulate a precise argument we need to know the $\lambda_k$ and $(\eta_k)_i$.
We have not managed to solve for the eigenvectors and eigenvalues of $M$ in general, but can do so for the toy model
we introduced in Section \ref{entgen}.
In this case $M$ has the following form
\bea
M=v^2 (1+{1\over p}){\bf 1}+v^2 K
\eea
where ${\bf 1}$ is the $p\times p$ identity matrix.
Everything is an eigenvector of the identity matrix, so we need only find the eigenvectors and eigenvalues of $K$.
Notice that $K$ has rank one.
Recall that the rank of $K$ is the dimension of the vector space spanned by its columns.
Since all of the columns of $K$ are identical they span a one dimensional space.
The rank is also equal to the number of non-zero eigenvalues, so $K$ has only one non-zero eigenvalue equal to $-1$.
The corresponding eigenvector $|k=-1\rangle$ has every component equal to 1
\bea
|k=-1\rangle={1\over\sqrt{p}}\left[
\begin{array}{c}
1\cr \vdots\cr 1
\end{array}
\right]
\eea
The remaining $K$ eigenvectors span the subspace orthogonal to $|k=-1\rangle$ and have $K$ eigenvalue equal to zero.
Thus, $|k=-1\rangle$ is an eigenvector of $M$ with eigenvalue equal to ${1\over p}$ while any vector orthogonal to 
$|k=-1\rangle$ is also an eigenvector of $M$ with eigenvalue equal to $1+{1\over p}$.
In this case the eigenvalues are not distinct and terms in (\ref{togetsmall}) will not in general cancel. 
Thus we don't expect $|\alpha|^2 |U_{ij}|^2$ to becomes small for large $t$ and hence the system will not equilibrate.
To get some insight into what is going on, consider an initial state, with excitations localized on a pair of lattice sites 
$i$ and $j$, given by
\bea
|\psi\rangle = {1\over\sqrt{2}}\left(b_i^\dagger -b_j^\dagger\right)|0\rangle
\eea
We easily find
\bea
H|\psi\rangle =\left(H_0+v^2+{v^2\over p}\right)|\psi\rangle +{v^2\over\sqrt{2}}
\sum_{k=1}^p (K_{ki}-K_{kj}) b_k^\dagger|0\rangle
\eea
The second term in the last line above allows excitations to move from their original lattice site to a new lattice site.
However the two contributions cancel so that $|\psi\rangle$ is an eigenstate and the excitation does not disperse - it 
remains localized on sites $i$ and $j$.
The excitation can move to every other site with exactly equal hopping strength, so that in the end the excitations
are blocked from moving anywhere and are instead localized.
This is rather generic: all states in the Hilbert space orthogonal to the state 
\bea
|k=-1\rangle ={1\over\sqrt{p}}\sum_{i=1}^p b_i^\dagger |0\rangle
\eea
are eigenstates of the Hamiltonian and hence do not evolve in time\footnote{At large $p$ the Hamiltonian becomes
a projector onto the space orthogonal to the $|k=-1\rangle$ so that the result of applying the Hamiltonian to any state
is an eigenstate.}.
The intuitive picture behind equilibration is as follows \cite{Calabrese:2006rx}: as time evolves, the system becomes 
correlated.
From each site a wave front moving at the speed of sound for the lattice emerges, carrying information.
The cumulative effect is an effective averaging process: information stored at one site becomes spread across
the entire lattice.
In our case, since the wave fronts are blocked from moving, we should not expect the system to equilibrate.

The spectrum of the typical Hamiltonian gives us an explanation for why we found such a small recurrence time.
In the large $p$ limit there is a single energy eigenvalue equal to ${v^2\over p}=0+O(p^{-1})$ and $p-1$ energy 
eigenvalues equal to $v^2-{v^2\over p}=v^2+O(p^{-1})$.
The energy of these degenerate states is the only energy in the problem and it clearly sets the recurrence time we found.
To get such a simple spectrum things must be fine tuned.
Random deviations from this typical Hamiltonian will lift the degeneracy leading to a spectrum that is more realistic
as we will soon see.

So, the Hamiltonian associated to the typical interaction graph exhibits localization.
However, even small fluctuations about this typical configuration should disrupt the localization.
Lets again look at a simple example. 
Choose a matrix $M_{ij}$ which opens up a ``conducting path'' that passes through each vertex of the graph, as follows
\bea
M=v^2(1+{1\over p}){\bf 1}+v^2K+v^2
\left[
\begin{array}{cccccc}
-{2\over p} &{1\over p}  &0                &\hdots &0                &{1\over p}\cr
{1\over p}  &-{2\over p} &{1\over p}   &\hdots &0               &0\cr
0               &{1\over p}   &-{2\over p} &\hdots &0               &0\cr
\vdots        &\vdots         &\vdots         &\ddots &\vdots         &\vdots\cr
0               &0                &0                &\hdots &-{2\over p} &{1\over p}\cr
{1\over p}  &0                &0                &\hdots &{1\over p}  &-{2\over p}
\end{array}
\right]
=v^2(1+{1\over p}){\bf 1}+v^2K+v^2L\nonumber
\eea
This is a small change to the typical Hamiltonian: we have only changed order $p$ matrix elements out
of a total of $p^2$ matrix elements, and we have only adjusted each element by an amount $\sim p^{-1}$.
The form of matrix $L$ was chosen so that we can again solve for the eigenvectors and eigenvalues of $M$ exactly.
First note that $[L,K]=0$ so that $L$ and $K$ can be simultaneously diagonalized.
A simple computation shows that the eigenvectors and eigenvalues of $L$ are given by 
$$
(\eta_k)_l = {e^{ikl}\over\sqrt{p}}\qquad \lambda_k={2\cos (k)-2+p-p\delta_{k,0}\over p}v^2 \qquad\qquad\qquad k={2n\pi\over p}
$$
with $n=0,1,2,\cdots,p-1$.
The eigenstate with $n=0$ is the eigenstate of $K$ with eigenvalue $-1$.
Using these eigenvalues and eigenvectors we have
\bea
U_{ij}&=&\sum_{n=0}^{p-1} e^{i{2\pi n\over p}(i-j)} e^{-i{2\cos ({2n\pi\over p})+p-p\delta_{k,0}-2\over p}v^2 t}
=U(i-j)\cr
&\sim& i^{i-j}e^{-iv^2}\times {1\over 2\pi i^{i-j}}\int_0^{2\pi} d\phi e^{-i{2\over p}\cos (\phi)v^2 t}
e^{i(i-j)\phi}\cr
&=& i^{i-j} J_{i-j}({2v^2 t\over p})
\eea
where $J_l(x)$ is the Bessel function.
This is the kind of result we want because we know that $|J_l(x)|<x^{-{1\over 3}}$ for all $x\ge 0$ \cite{bessel}.
Thus, for times such that
\bea
{2v^2 t_{\rm eq}\over p}\sim 1\qquad\Rightarrow\qquad t_{\rm eq}\sim {p\over 2v^2}
\eea
the matrix elements $U_{ij}$ are becoming small enough to neglect and the density matrix is approaching a Gaussian state.
Thus, the system evolves to the state of maximum entropy and we come to equilibrium.
Notice that this time is much much smaller than the enormous recurrence time.
The system now remains at equilibrium until we get close to the recurrence time.
Notice that $t_{\rm eq}$ is significantly larger than the scrambling time.

\section{Discussion}\label{discuss}

We have studied the one loop mixing problem for operators with a large enough bare dimension that they
could be dual to black holes or new spacetime geometries.
This mixing problem is significantly more complicated than the planar mixing problem.
Despite this, a remarkably simple description emerges.
The dilatation operator defines dynamics on a graph of the type that has recently been suggested as models
for quantum dynamics of black holes \cite{Lashkari:2011yi,Bentsen:2018uph,Lucas:2019aif,Chen:2019klo,Garcia-Garcia:2020cdo,Xu:2020shn}.
It is intriguing to see simple dynamics on graphs emerging naturally from the mixing problem of very large dimension
operators in Yang-Mills theory.

Each operator has a number of labels, one of which is the interaction graph.
Operators only mix if they have the same interaction graph label.
We have carried out a careful counting of the interaction graphs and find that the number of graphs matches the entropy
of a black hole suggesting that we might think of these operators as dual to a black hole microstate.
By numerically generating lists of graphs we have characterized the ``typical interaction graph'' and the dynamics
associated to it.
We find a lattice model defined on $p$ sites with $p\sim O(N)$ and with all-to-all interactions.
Despite this non-locality, we have proved that the scrambling time is bounded consistent with the fast scrambling conjecture.
By considering a specific example, we have also given evidence that the system equilibrates in a time scale
$t\sim {p\over\lambda}$ where $p\sim N$.

The idea that gravitational dynamics should emerge from the sector of heavy operators in the Yang-Mills theory has been
pursued in \cite{Berenstein:2005aa,Koch:2008ah,Koch:2009gq,Berenstein:2014pma,Lin:2017dnz,Berenstein:2017rrx}.
Our study is a continuation of these ideas.

There are a number of interesting directions that could now be pursued.
Our analysis has all been limited to weak coupling.
To make contact with black hole physics we need to make progress in understanding the strong coupling limit
of the theory, which is presently a formidable problem.
However, one might look for BMN like \cite{Berenstein:2002jq} limits or for observables that are protected by super 
symmetry, which has not yet been considered in the setting of heavy operators.
A more manageable problem is to generalize our analysis to generic operators constructed using all of the fields
in the field theory.
By using only complex scalar fields $\phi_i$ and not $\phi_i^\dagger$, we naturally construct operators that have
dimension close to their $R$-charge.
By including enough $\phi_i^\dagger$ fields we would be able to construct operators with the 
quantum numbers expected for near extremal or even Schwarzschild black holes.
This generalization should be a straight forward technical exercise.
The spectrum we have computed may find application in the arguments of \cite{Berenstein:2018hpl}
which explore how the thermodynamics of small black holes is recovered from the dual conformal field theory.
Our considerations of equilibration made use of two specific examples and a specific initial condition.
Clearly a lot more is needed to properly understand the equilibration of our system and the associated time scales.
It would also be interesting to explore situations in which we need to correct the distant corners approximation, which
are required when the giant gravitons become coincident in space time.
Thermal averages in the Yang-Mills theory involve averages over the complete ensemble of graphs.
Corrections to the distant corners approximation would allow transitions between different graphs and the
number of particles hopping on the graph would no longer be conserved.

Finally, our goal was to gain some insights into the mechanism behind extremely rapid black hole thermalization rates 
which must be present in the dynamics of large $N$ Yang-Mills theories.
Since our study has reduced to simple dynamics on graphs, perhaps the most important lesson to be drawn is that the 
``toy models'' considered in 
\cite{Lashkari:2011yi,Bentsen:2018uph,Lucas:2019aif,Chen:2019klo,Garcia-Garcia:2020cdo,Xu:2020shn} may in 
fact be better than one might have expected.
The description in terms of a graph certainly carries over to the case that more fields are included, but its validity
at strong coupling is yet to be established.

{\vskip 0.4cm}
\noindent
\begin{centerline} 
{\bf Acknowledgements}
\end{centerline} 

This work is supported by the Science and Technology Program of Guangzhou (No. 2019050001), by a Simons Foundation 
Grant Award ID 509116 and by the South African Research Chairs initiative of the Department of Science and Technology 
and the National Research Foundation.
We are grateful for useful discussions to Sanjaye Ramgoolam. 

\appendix

\section{Gauss Graph Hamiltonian from Yang-Mills}\label{fromD}

In this section we review the results of \cite{deCarvalho:2018xwx,deMelloKoch:2020agz}, where the Hamiltonian we 
study (\ref{GGham}) was derived.
We consider the mixing problem for operators belonging to the su$(2|3)$ sector of the theory.
Truncation to this subsector is consistent to all orders of perturbation theory\cite{Beisert:2003jj}. 
We choose this sector because it is the maximal closed subsector with finitely many fields.
The fact there are finitely many fields simplifies the analysis and it is possible to obtain explicit formulas for the action
of the dilatation operator.

A basis for these operators is given by the restricted Schur polynomials.
The relevant restricted Schur polynomials are labeled by 6 Young diagrams and some multiplicity labels.
We study operators with $\Delta\sim N^2$ that are holographically dual to a system of giant gravitons. 
Operators with $p$ long columns (rows) are dual to a system of $p$ (dual) giant gravitons\footnote{Branes 
connected by an open string described using a spin chain have been considered in
\cite{Berenstein:2013md,Berenstein:2013eya,Berenstein:2014isa,Berenstein:2014zxa,Koch:2015pga}.}.
These operators mix with each other, but not with operators labeled by Young diagrams of a different shape. 
We take $n_1\sim \epsilon N$ with $\epsilon\ll 1$.
There are bosonic $\phi_2,\phi_3$ excitations, as well fermionic ($\psi_1$ and $\psi_2$) excitations.
Limit the number of excitations by requiring $n_2\sim n_3\sim m_1\sim m_2\sim \epsilon^2 N^2$.
We use a collective label $N_A=(n_2,n_3,m_2,m_2)$ to refer to the number of excitations.

We will now explain why our operators are labeled by Young diagrams.
To construct all possible gauge invariant operators, we can take a product of an arbitrary number of fields and then
contract all row indices with all column indices to obtain a gauge invariant operator.
We can specify which row indices are to be computed with which column indices by giving a permutation.
So we could label our operators with a permutation.
Alternatively, by taking a Fourier transform on the group, we can trade the permutation for the label of
an irreducible representation, that is, for a Young diagram\footnote{This discussion is not quite the whole story.
We should have one Young diagram for the row indices and one for the column indices. Projecting to the
singlet then forces these two to agree.}.  
This introduces the Young diagram $R$ which has as many boxes as fields used to construct the operator, i.e. it has
$n_1+n_2+n_3+m_1+m_2$ boxes. 
For operators dual to giant gravitons\cite{McGreevy:2000cw}, the Young diagram $R$ has a small number of long columns 
and for operators dual to dual giant gravitons\cite{Grisaru:2000zn,Hashimoto:2000zp}, the Young diagram $R$ has 
a small number of long rows\cite{Balasubramanian:2001nh,Corley:2001zk,David}.
We will consider operators with a total of $p$ long rows.

The construction described so far is redundant.
Distinct permutations used to construct the gauge invariant operator might only differ by swapping two fields of a given
species.
Since our fields are bosons or fermions, swapping these fields does not lead to a new operator.
Consequently, we need to remove this redundancy.
This is done by projecting so that the collection of fields of a given species is in a definite representation of the
permutation group - so we get one more Young diagram for each species of field\footnote{Recall that the permutation
group swapping indices of all fields has appeared. $R$ is a representation of this group. The representations for each 
species are a representation of the subgroup which swaps only indices of fields that are the same species. The representation
of the subgroup can be embedded into $R$ in more than one way and this is why we need multiplicity labels.}.
Since we have five different types of fields, this makes a total of 6 Young diagrams.  
Each box in the Young diagram $R$ corresponds to a field, and we can specify how many fields of each species appear
in a given row of $R$.
This specifies the excitations of each dual giant graviton brane.

The operators that are obtained by this construction have orthogonal two point functions in the free field 
theory\cite{Bhattacharyya:2008rb}, provide a complete linear basis for local gauge invariant 
operators \cite{Bhattacharyya:2008xy} and they mix only weakly when interactions are turned on \cite{Koch:2011hb}.
The Hamiltonian we study is derived by evaluating the action of the one loop dilatation operator in the su$(2|3)$ sector,
which is given by \cite{Beisert:2003ys,Eden:2004ua}
\bea
  D=&-&{2g_{YM}^2\over (4\pi)^2}\left(\sum_{i>j=1}^3 \, {\rm Tr}\left(\left[\phi_i,\phi_j\right]\left[\partial_{\phi_i},\partial_{\phi_j}\right]\right)
     +\sum_{i=1}^3\sum_{a=1}^2 \, {\rm Tr}\left(\left[\phi_i,\psi_a\right]\left[\partial_{\phi_i},\partial_{\psi_a}\right]\right)\right.
\cr
&+&\,{\rm Tr}\left(\left\{\psi_1,\psi_2\right\}\left\{\partial_{\psi_1},\partial_{\psi_2}\right\}\right)
\Bigg)
\label{fullD}
\eea
on restricted Schur polynomials.
It is useful to introduce the notation
\begin{eqnarray}
D\equiv -{2g_{YM}^2\over (4\pi)^2}\sum_{A>B=1}^5 D_{AB}
\end{eqnarray}
where $D_{AB}$ mixes fields of species $A$ and $B$.
A major simplification in this computation follows by noting that at large $N$, corners on the right hand side of the Young
diagram are well separated.
This is the displaced corners limit \cite{Carlson:2011hy,Koch:2011hb}.
The action of the symmetric group simplifies in this limit and there are new symmetries: swapping the row or column indices
of fields that belong to a given species and sit in the same row of $R$ is a symmetry.
To use these new symmetries we refine the number of fields of a species $N_A$ to produce a $p$ dimensional vector 
$\vec N_A$, with each component recording how many fields are in a given row.
For example, the number of $\phi_2$ fields $n_2$ is refined to produce $\vec n_2$, and the group swapping $\phi_2$ fields 
in a given row, the enhanced symmetry of the displaced corners limit, is\footnote{We divide on the left to account for the
symmetry associated with the row indices and on the right to account for the symmetry associated with column indices.
See (\ref{dcosets}).}
\begin{equation}
   H_{\vec{n}_2}=S_{(n_2)_1}\times S_{(n_2)_2}\times \cdots\times S_{(n_2)_p}
\end{equation}
In this limit, the number of restricted Schur polynomials matches the order of the double coset, indicating that we can 
organize the local operators using the double coset \cite{deMelloKoch:2012ck}.
The four double cosets relevant for labeling our operators are
\begin{eqnarray}
A&\leftrightarrow& \sigma_{A}\in H_{\vec{N}_{A}}\setminus S_{N_{A}}/H_{\vec{N}_{A}}
\label{dcosets}
\end{eqnarray}
These double cosets are the crucial ingredient needed to make the connection to physics on a graph.
Indeed, the collection of graphs with $n$ edges and $p$ vertices, and with number of edges terminating at each vertex 
recorded in $\vec n$ is described by a double coset \cite{deMelloKoch:2011uq}.
By this connection each element of a double coset is described by a graph, so that we can label our operators by a graph. 
Diagonalizing $D_{\phi_1,A}\in\{D_{\phi_1\phi_2},D_{\phi_1\phi_3},D_{\phi_1\psi_1},D_{\phi_1\psi_2}\}$ first,
the resulting eigenoperators are the Gauss graph operators \cite{Koch:2011hb,deMelloKoch:2012ck}, labeled by two 
Young diagrams (the $R$ and $r_1$ labels of the restricted Schur polynomial) and a graph (which takes the place of
four Young diagrams).
Vertices of graphs correspond to rows/columns of $r_1$, i.e. each vertex corresponds to a giant graviton brane.
Each $A$ field type is a species of edge in the graph and there is an edge for each field.
Edges are directed.
We give the complete graph as a graph for each $A$, specified by four elements $\vec\sigma$, one of each of the 
four double cosets in (\ref{dcosets}).
These are the graphs that we call Gauss graphs.

Intuitively its clear why the graph provides a useful description: it naturally accounts for the symmetries of the
displaced corners limit.
Recall that each row of $R$ corresponds to a vertex and each edge in the graph corresponds to a field in the operator.
The symmetry of swapping row indices of fields in a given row is now the symmetry of swapping endpoints of edges
that end on the same vertex (an obvious symmetry of the graph) while the symmetry of swapping column indices of 
fields is the symmetry of swapping start points of edges that start on the same vertex.

The elements of the double cosets in (\ref{dcosets}) correspond to the graphs we consider.
Vertices can be dressed by closed edges with ends attached to the same vertex or by edges between two distinct vertices.
Fermi statistics forbids two or more parallel edges (edges with the same orientation and endpoints) of the same 
fermion species \cite{deCarvalho:2020pdp}.
We refined $N_A$ to produce a vector $\vec N_A$.
To describe the graph refine $\vec N_A$ to produce a matrix $(N_A)_{i\to j}$ whose elements 
describe the number of edges running from vertex $i$ to vertex $j$.
In terms of this matrix, the Gauss Law constraint is $\sum_{k\ne i} (N_A)_{i\to k}=\sum_{k\ne i} (N_A)_{k\to i}$.
The transformation from restricted Schur basis to Gauss graph basis is derived in \cite{Koch:2012sf}.
After the transformation, the dilatation operator is most naturally written as a system of particles hopping on a lattice,
with lattice sites given by vertices of the Gauss graph \cite{deCarvalho:2018xwx}.
Closed edges forming loops at a vertex translate into particles at that site.
The hopping strength is determined by the number of edges of all other species stretched between the vertices.
There are two distinct species of bosons, for $\phi_2,\phi_3$, and two distinct species of fermions, for $\psi_1,\psi_2$.

To simplify the discussion that follows, we will consider only the bosonic sector of the theory.
The bosons are described by oscillators
\begin{eqnarray}
\big[ a_{ij},\bar a_{kl}\big]=\delta_{il}\delta_{jk}\qquad\qquad
\big[ b_{ij},\bar b_{kl}\big]=\delta_{il}\delta_{jk}
\end{eqnarray}
with all other commutators vanishing.
The Fock space vacuum $|0\rangle$ obeys $a_{ij}|0\rangle =0=b_{ij}|0\rangle$ for $i,j = 1,2,\cdots ,p$.
our final result for the Hamiltonian of the lattice model, arising from the one loop dilation operator, is
\bea
H&=&{2g_{YM}^2\over (4\pi)^2}\sum_{A=1}^4\sum_{i>j=1}^p(\hat{N}_A)_{ij}
\left(\sqrt{N+l_{R_i}}-\sqrt{N+l_{R_j}}\right)^2\cr
&&+{2g_{YM}^2\over (4\pi)^2}\sum_{A=1}^3\sum_{B=1+A}^4
\sum_{i,j=1}^p\sqrt{(N+l_{R_i})(N+l_{R_j})\over l_{R_i}l_{R_j}}\Bigg(
-(\hat{N}_B)_{ji}(\bar{a}_{A})_{jj} (a_{A})_{ii} 
-(\hat{N}_A)_{ji}(\bar{a}_{B})_{jj} (a_{B})_{ii}\cr\cr
&&\quad +2\delta_{ij} 
\Big(\sum_{l\ne i} (\hat{N}_A)_{i\to l} + (\bar{a}_A)_{ii}(a_A)_{ii}\Big)
\Big(\sum_{l\ne i} (\hat{N}_B)_{i\to l} + (\bar{a}_B)_{ii}(a_B)_{ii}\Big)
\Bigg)
\label{LatticeH}
\eea
In the above formula, $l_{R_i}$ is the length of the i$th$ row of Young diagram $R$.

Thus, in this non-planar limit the operator mixing problem translates into dynamics on an emergent lattice,
described by a graph.
We consider states with definite $(\hat N_A)_{i\to j},(\hat N_A)_{ij}$ for $i\ne j$ eigenvalues.
These are constants of the motion.
We can replace the operators $(\hat N_A)_{i\to j},(\hat N_A)_{ij}$, for $i\ne j$ by fixed non-negative integers 
$(N_A)_{i\to j},(N_A)_{ij}$ for each state.
To simplify the problem we consider operators without fermionic excitations
\bea
(\hat{m}_1)_{ij}=0=(\hat{m}_2)_{ij}
\eea
Edges between vertices are given by the $\phi_2$ field, so that
\bea
(\hat{n}_3)_{ij}=0\qquad i\ne j
\eea
Excitations localized to a vertex are all given by $\phi_3$ fields so that
\bea
(\hat{n}_2)_{ii}=0
\eea
Setting $(n_2)_{ij}=N_{ij}$, $r_i=N+l_{R_i}$, $k_i=\sum_{j\ne i,j=1}^p N_{ij}$ and renaming $l_{R_i}\to l_i$ 
$b_{ii}\to b_i$ and $\bar{b}_{ii}\to b_i^\dagger$ our Hamiltonian becomes
\bea
H&=&{g_{YM}^2\over (4\pi)^2}\sum_{i,j=1}^p N_{ij}
\left(\sqrt{r_i}-\sqrt{r_j}\right)^2
-{2g_{YM}^2\over (4\pi)^2}\sum_{i,j=1,i\ne j}^p\sqrt{r_i r_j\over l_{i}l_{j}}N_{ji}b_{j}^\dagger b_i
+{2g_{YM}^2\over (4\pi)^2}\sum_{i=1}^p
{k_i r_i\over l_i}b_i^\dagger b_i\cr
&&
\eea
This is the one loop correction to the dimension.
To get the total dimension of the operator we would sum this with the bare dimension, given by
$\Delta_0=n_1+n_2+n_3+{3\over 2}(m_1+m_2)$.

\end{document}